\documentclass[prb,superscriptaddress,a4paper,sort]{revtex4}
\usepackage{amssymb}
\usepackage{citesort,epsfig}
\begin{document}

\title{The effects of added salt on the second virial coefficients
of the complete proteome of {\em E. coli}}
\author{Richard P. Sear}
\affiliation{
Department of Physics, University of Surrey, Guildford,
Surrey GU2 7XH, United Kingdom\\{\tt email: r.sear@surrey.ac.uk}}

\begin{abstract}
Bacteria typically have a few thousand different proteins.
The number of proteins with a given charge is a roughly
Gaussian function of charge --- centered near zero,
and with a width around ten (in units of the charge
on the proton). We have used the charges on 
E. coli's proteins to
estimate the changes in the
second virial coefficients of all its proteins as the
concentration of a 1:1 salt is increased.
The second virial
coefficient has dimensions of volume and we find that on average it decreases
by about twice the average volume of a protein
when the salt concentration is increased from $0.2$ to 1 Molar.
The standard
deviation of the decrease is of the same order.
The consequences of this for the complex mixture of proteins inside
an E. coli cell, are briefly discussed.
\end{abstract}

\maketitle

\section{Introduction}

The genomes of a number of organisms are already known and more are
being completed at a rate of perhaps one a month.
Once a genome \cite{nomen}
has been sequenced the amino-acid sequences of {\em all} its
proteins are known. The complete set of proteins
of an organism is generally referred to as its proteome. 
Here we use genome data for {\em E. coli} to systematically
estimate the change in the interactions of all the proteins of this
bacterium when the salt concentration is varied.
{\em E. coli} can grow in environments with a very
wide range of salt concentrations \cite{cayley91}, and so its
proteins must function {\em in vivo} over a wide range of salt
concentrations. The 
potassium ion concentration inside the cell can vary from approximately
$150$ to $300$ mM \cite{cayley91};
potassium is the predominant cation in living cells.
Clearly, the proteins must remain soluble over this
range, and they should bind to the other proteins which they are required
to bind to in order to function, but they should not
interact strongly with other proteins.
The study of a proteome is
often called proteomics. Here, as the physical properties of proteins
are studied (as opposed to their chemical properties such as catalytic
function) we are doing what may be called physical proteomics.

There has been extensive theoretical work
on the salt dependence of the interactions in individual proteins,
particularly for the protein lysozyme
\cite{poon00,warren02,muschol97}.
See
Refs.~\onlinecite{poon00,muschol97,velev98,guo99,rosenbaum99,piazza00}
for corresponding experimental work.
However, as far as the author is aware,
this is the first attempt to characterise the interactions of
{\em all} the proteins of an organism. We will consider the proteins
separately, i.e., as single component solutions. Of course inside
a bacterium the proteins exist as a mixture of thousands of components.
Future work will consider multi-component mixtures of the proteins.
We have chosen {\em E. coli} as
it is a bacterium, and therefore a relatively simple organism, and as
it has been extensively studied. However, the distribution of
charges on the proteins of almost all organisms is very similar
and so our results apply to almost all organisms, including
{\em H. sapiens}. The only exceptions are some extremophiles
\cite{runcong01}.

In the next section we use genome data to estimate the charges on the
proteins of {\em E. coli}. This data is used in the third section where
we calculate the variation in their second virial coefficients
as the salt concentration is varied. The last section is a conclusion.

\section{The charges on proteins of E. coli}

{\em E. coli K-12} has a proteome of 4358
proteins. The amino acid sequences of all of them are known
from the sequencing of its genome \cite{coligen,ebi}.
{\em K-12} is the name of a strain of {\em E. coli}.
Runcong and Mitaku have analysed the charge distributions
of a number of organisms using
a simple approximate method of estimating the charge on a protein at
neutral pH from its amino-acid sequence \cite{runcong01}.
We will follow their analysis but use a slightly different approximation
for the charge on a protein with a given amino-acid sequence.

Of the 20 amino acids, 5 have pK values such that they should be at least
partially charged at neutral pH \cite{thecell,stryer}. These
are two highly acidic amino acids, aspartic acid and glutamic acid,
two highly basic amino acids, lysine and arginine and one somewhat basic
amino acid, histidine.
Aspartic and
glutamic acids have pK's far below 7 and lysine and arginine have pK's
far above 7 and so we assume that all 4 of these amino acids are fully
charged at neutral pH. Aspartic and
glutamic acids then each contribute $-1$ to the charge on a protein,
and lysine and arginine each contribute $+1$.
Histidine has a pK of around 6-6.5 (this will depend
on the environment of the amino acid). The equation for the fraction $f$ of
a basic group such as histidine that is charged at a given pH is
\begin{equation}
f=1/\left(1+10^{pH-pK}\right),
\end{equation}
where pK is the pK value for the basic group. This equation
is just the Henderson-Hasselbalch equation \cite{stryer} rearranged.
Taking pK$=6.5$ \cite{stryer} and at
pH=7 we have that the fraction of histidines charged is
$f=0.24$. As this is small we assume for simplicity that all the
histidine amino acids are uncharged.

Thus, with these assumptions for the charges on these
5 amino acids, our estimate for the
net charge $Q$ on a protein is simply given by
\begin{equation}
Q=n_K+n_R-n_D-n_E
\label{qdef}
\end{equation}
where $n_K$, $n_R$, $n_D$ and $n_E$ are the the protein's total numbers of
lysines, arginines, aspartic acids and glutamic acids, respectively. The
subscripts $K$, $R$ etc. correspond to the standard single letter
codes for the amino acids \cite{stryer,thecell}. The charge $Q$ is
in units of $e$ where $e$ is the elementary charge.

Note that Runcong and Mitaku \cite{runcong01} assume that the histidine
amino acids contribute $+1$ to the charge,
that is the only difference between our analysis and that of
Runcong and Mitaku. As the histidine amino acid is quite a rare amino acid,
approximately 1 in 50 amino acids is a histidine, the difference
between the results we obtain and those of Runcong and Mitaku \cite{runcong01}
is not large but our charges are shifted to more negative values.
Using Runcong and Mitaku's approximation the mean charge on a protein
is $7.11$ units more positive than the mean charge we find here.
As a check on our algorithm, we can compare the prediction of
equation (\ref{qdef}) for chicken lysozyme
to that of a titration experiment to determine the charge.
Equation (\ref{qdef}) predicts that chicken lysozyme \cite{aalys}
has a net charge
of 8 at neutral pH. Titration experiments on lysozyme
give a titratable charge of close to $8.5$ at pH=7 \cite{lysoq}.

\begin{figure}
\begin{center}
\vspace*{0.3in}
\epsfig{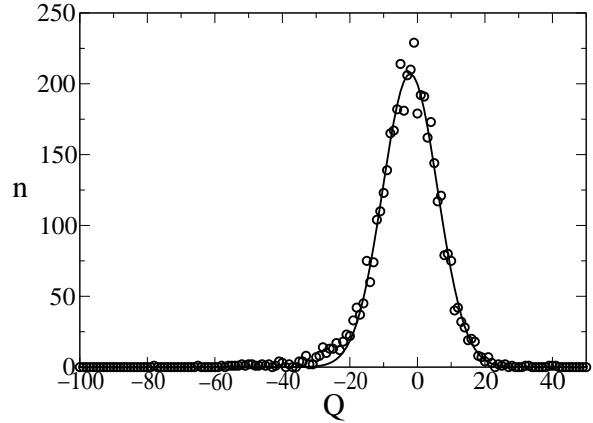}
\end{center}
\caption{
The number of {\it E. coli K-12} proteins $n$ as a function
of net charge $Q$. There are 2 proteins with net charges $<-100$;
these are not shown. The data for each value of $Q$ are shown
as circles and the curve is a Gaussian fitted to the data.
}
\label{qcoli}
\end{figure}

Using equation (\ref{qdef}) we can obtain estimates for the charges of
all 4358 proteins of {\em E. coli}
\cite{coligen,ebi,gerstein98,mitaku99}.
The results are shown
in Fig.~\ref{qcoli}, where we have plotted the number of proteins
$n$ as a function of net charge $Q$.
The distribution is centered almost at a net
charge $Q=0$, and for not-too-large $|Q|$ the distribution
is roughly symmetric and Gaussian. The mean charge is $-3.15$. Given
the approximate nature of our equation for the charge on a protein,
equation (\ref{qdef}), the data
is probably consistent with a mean charge of 0.
The approximation scheme of Mitaku and
Runcong \cite{runcong01} yields a mean charge of $+3.96$.
Also, although when $|Q|$ is not too large the
distribution can be seen to be reasonably symmetric, {\em E. coli}
has 12 proteins with charges $<-50$ but none with charges $>+50$.
Excluding proteins with very large charges, $|Q|>30$, the root
mean square charge equals $9.1$.

A number of other organisms,
both other bacteria and eukaryotes such as yeast, have had
the charge distribution on their proteomes determined
by Runcong and Mitaku \cite{runcong01} and by the author
\cite{unpub}. Almost all of them have a roughly Gaussian
distribution centered approximately at zero, like the distribution
in Fig.~\ref{qcoli}. The exceptions are some extremophiles.
Extremophiles are organisms
that live in extreme environments, for example
{\em Halobacterium sp.} lives in environments with very high
levels of salt \cite{microbio}. The cytosol of {\em Halobacterium sp.}
contains much higher levels of potassium ions
than do other organisms so perhaps it is not a surprise that the
distribution of charges on its proteins is different.

We have fitted the Gaussian function
\begin{equation}
n(Q)=\frac{1739}{\sigma}\exp(-(Q-{\overline Q})^2/2\sigma^2).
\label{pq}
\end{equation}
to the data for the number of proteins
as a function of their charge. It is drawn as the solid curve
in Fig.~\ref{qcoli}. The fit parameters are
mean charge ${\overline Q}=-2.16$ and standard deviation
$\sigma=8.32$. 1739 is $4358/(2\pi)^{1/2}$ and so the distribution
is normalised so that its integral gives the total number of proteins.
Within a couple of standard deviations of the mean the
Gaussian function fits the data well but it underestimates the numbers
of proteins with charges such that $|Q-{\overline Q}|$ is
several times the standard deviation.

We also note that there is a correlation between the net charge
$Q$ on a protein and its size, measured by the number of amino acids $M$.
Figure \ref{scat_coli} is a scatter plot of charge $Q$ and number
of amino acids $M$ for the proteins of {\em E. coli}. Although
at any particular size $M$ there is a wide distribution of charges,
on average the more highly charged proteins are larger
than average.
We expect
the volume of a protein to scale with $M$.

\section{Salt dependence of the second virial coefficients}

Consider a dilute solution of a single one of the proteins of
{\em E. coli}.
Apart from water, the only other constituents are
a 1:1 salt at a concentration $c_s$ and a buffer which
controls the pH while making a negligible contribution to the ionic strength.
Here we will always assume the pH=7 but
other pH's can be considered if the net charges on the proteins
can be calculated. Also, the counterions of the protein
are assumed the same as either the anions or cations of the salt, depending
on the sign of $Q$.
The interactions between
the protein molecules in the salt solution can be characterised
by means of the protein's second virial coefficient $B_2$:
a function of temperature, pH and salt concentration.

Proteins are complex molecules and we are unable to calculate
from first principles the absolute value of $B_2$ for any of the
4358 proteins possessed by {\em E. coli}. However, predicting
the change in the second virial coefficient when the salt concentration
varies
is a much easier problem, {\em if} we assume that changing the salt
concentration changes only the direct electrostatic interaction
between the net charges of a protein. This is a strong assumption
but studies of the simple protein lysozyme have shown that the
variation of its second virial coefficient can be described using
a simple model which only includes its net
charge \cite{poon00,warren02}. Here we will follow
Warren \cite{warren02} and apply his analysis of lysozyme
to the complete set of proteins of {\em E. coli}. We will discuss
which proteins are likely to be less well described by this theory
than is lysozyme.

A protein molecule of charge $Q$ is surrounded by its counterions
and as the concentration of the protein is increased so is
the counterion density. This increase in the counterion density
decreases the translational entropy of the counterions and this
contributes a positive amount to the second virial coefficient.
See Warren \cite{warren02} and references therein for details.
$B_2$ has the form \cite{warren02}
\begin{equation}
B_2=B_2^{(ne)}+\frac{Q^2}{4c_s},
\label{donnan}
\end{equation}
where $B_2^{(ne)}$ is an assumed constant term due to excluded volume
interactions and other interactions which are insensitive to salt
concentration. The second term is from the counterions and the salt.
It is quadratic in the charge and so of course is zero for uncharged
proteins and is independent of the sign of the net charge on a protein.

\begin{figure}
\begin{center}
\vspace*{0.3in}
\epsfig{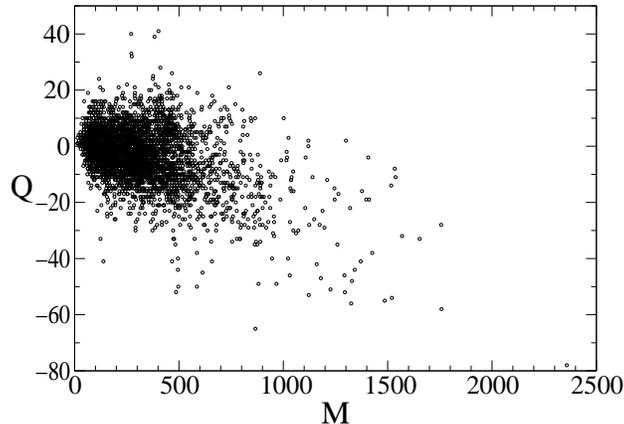}
\end{center}
\caption{
A scatter plot of the charge on a protein $Q$ versus the number
of amino acids $M$. All but 2, both with charges $<-100$,
of the proteins of {\it E. coli K-12} are shown.
}
\label{scat_coli}
\end{figure}

\begin{figure}
\begin{center}
\vspace*{0.3in}
\epsfig{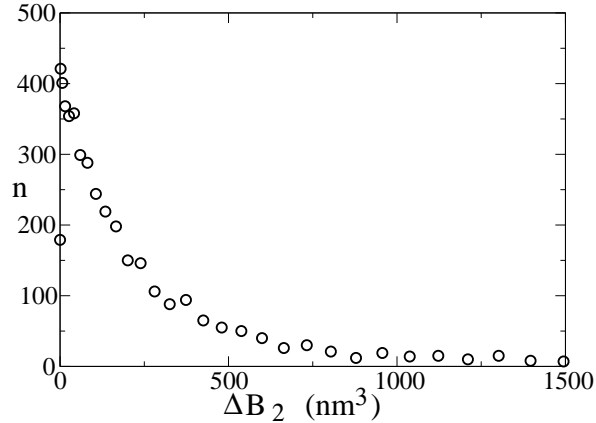}
\end{center}
\caption{
The number of {\em E. coli K-12} proteins $n$ as a function
of the change
in their second virial coefficient, $\Delta B_2$, when the salt
concentration is decreased from 1 Molar to 0.2 Molar.
Results are only shown for proteins with $|Q|\le30$.}
\label{db2}
\end{figure}

As stated above we are unable to calculate the absolute value
of $B_2$; the values of $B_2^{(ne)}$ of the proteins are unknown.
However, we can calculate the difference in $B_2$ when the salt
concentration is changed from $c_s=c_1$ to $c_s=c_2$. It is
\begin{equation}
\Delta B_2=\frac{Q^2}{4}\left(\frac{1}{c_2}-\frac{1}{c_1}\right).
\label{dond}
\end{equation}
This is easy to calculate for any protein and in
Fig.~\ref{db2} we have plotted
the number of proteins $n$ as a function of the change in
their second virial coefficient, $\Delta B_2$, when the
salt concentration is decreased from 1M to $0.2$M.
The results are given in units of nm$^3$. For comparison the
volume of a typical bacterial protein is about 60nm$^3$ and so
if a protein were to interact solely via a hard repulsion it
would have a second virial coefficient of about 4 times its volume
or about 240nm$^3$.

Results for proteins with $-30\le Q\le30$ are shown. Proteins
with larger titratable charges are likely to have an effective
charge lower than $Q$, see Refs.~\onlinecite{lobaskin01,warren00}
and references therein.
From the linear Poisson-Boltzmann equation, the potential (divided by $e$)
at the surface of a spherical
particle with charge $Q$ and radius
$a$ is $Q\lambda_BkT/((1+\kappa a)a)$.
$\lambda_B=e^2/(4\pi\epsilon kT)$ is the Bjerrum length,
and and $\kappa^{-1}$ is the Debye screening length, given by
$\kappa^2=8\pi\lambda_Bc_s$.
For the dielectric constant of water 80 times that in vacuum and
at room temperature, $\lambda_B=0.7$nm.
Globular proteins are approximately spherical and typically have radii
around 2 to 4nm. Taking a protein with a radius of 3nm,
in salt at a concentration $c_s=0.1$M,
we have that for $Q=30$, the potential
at the surface is about $2kT$.
Larger charges correspond to
larger surface potentials and these large potentials bind oppositely
charged ions to the surface reducing the effective charge.
On average, this effect will be diminished to a certain extent by the fact that
the most highly charged proteins are larger than average.
See Fig.~\ref{scat_coli}, where it is clear that the charge and
size of a protein are correlated.
Recent simulations by Lobaskin {\it et al.} \cite{lobaskin01}
of spheres with radius 2nm and charge $Q=-60$
in the absence of salt found an effective charge of a little under $-20$.
Thus we restrict ourselves to proteins with charges
of magnitude less than or equal to 30. 4300 of the 4358 proteins, or
almost 99\%, have charges in this range.
The mean change in $B_2$ of these 4300 proteins when the salt
concentration is decreased from 1 to $0.2$M is 139nm$^3$ and the
standard deviation is 218nm$^3$.

A couple of caveats. The first is that the effect of salt on protein
solutions is known to depend not only on whether the salt is a 1:1 salt, a
1:2 salt etc. but also to the nature of ions, whether it is Mg$^{2+}$
or Ca$^{2+}$ for example \cite{durbin96}.
Our generic theory applies only where there are no
specific interactions between the salt and the protein. There is
good agreement between experiment and
theory for lysozyme plus NaCl \cite{warren02,poon00} and so we may
hope that it applies to NaCl and many proteins but it clearly
misses potentially important effects for other salts where there
are specific protein-salt interactions. The second is that proteins
are not simple charged spheres, for example some have large dipole
moments. Dipoles exert net attractions which are screened and hence
weakened by added salt. Thus proteins with a small charges but large
dipole moments are poorly described by the current theory: if the
dipole interactions are dominant then the second virial coefficient
may even increase when the salt concentration is increased.
Velev {\em et al.} \cite{velev98} discuss this point.
Note that although we can estimate the charge on a protein from its
amino-acid sequence we cannot estimate its dipole moment without
knowing its three-dimensional structure, and so the sequence
data from genomics is not adequate to determine dipole moments.

\section{Conclusion}

Here we have
shown how data from genomics can be used to estimate the charges
on the proteins of an organism. We then used these charges to
estimate the changes in the second virial coefficients of 4300
(99\%) of the proteins of {\em E. coli}
when the salt concentration is changed. Note that {\em E. coli} can
survive and multiply in external environments with a very
wide range of salt concentrations; Cayley {\em et al.} \cite{cayley91}
studied the growth of {\em E. coli} in environments with
salt concentrations ranging from
very low to $0.5$ Molar, corresponding to
potassium ion concentrations inside the cell of
$150$ to $300$ mM \cite{cayley91}.
Thus, studying the change in interactions of
proteins with salt concentration is of direct relevance to the
{\em in vivo} behaviour of proteins.
Within molecular biology there is a clear shift of emphasis
away from studying the proteins of an organism one or a few at a time, and
towards determining the structure and function of large sets of proteins,
in particular proteomes.
The systematic study of these large sets of proteins is often
called proteomics.
This work is a first attempt to keep up with this shift by performing
a simple theoretical calculation of a solution phase physical property
for a complete proteome, rather than for one or a handful of proteins
as is usually done. It may be termed physical proteomics.

\begin{table}
\begin{tabular}{|c|c|c|c|}
\hline
Charged species & $Q$ & n & Charge density in cytosol (mM) \\
\hline
Protein & $-21$ to $+15$ & $2\times10^6$ & $-10$ \\
tRNA & $-80$ & $2\times10^5$ & $-30$ \\
mRNA & $-2000$ & $1\times10^3$ & $-3$ \\
Ribosome & $-3000$ & $2\times10^4$ & $-100$ \\
DNA & $-10^{7}$ & 1 & $-20$ \\
\hline
\end{tabular}
\caption{The charged macromolecular species in the cytosol of {\em E. coli}.
The data is from Neidhardt \cite{neidhardt}.
$Q$ is the charge on a macromolecule. For proteins the range given
is the mean plus and minus twice the standard deviation.
$n$ is
the total number of molecules of a species per cell.
The volume of an {\em E. coli} cell is about $10^{-18}$m$^3$
so 1 molecule per cell corresponds to a concentration of about
$2\times 10^{-9}$ Molar.
A prokaryote ribosome consists of about
4500 bases of RNA plus protein. The ribosomal proteins are mostly
quite stongly positively charged and so will decrease
the net negative charge. As a rough estimate we settle on a net charge of
$-3000$.
The charge on a tRNA molecule is
around $-80$, $-1$ from each of its bases \cite{stryer,thecell}.
The charge on a mRNA molecule is on average
around $-2000$ \cite{neidhardt}.
The charge on DNA is equal to
twice the number of base pairs, 4,639,221 for {\em E. coli K-12}
\cite{coligen}. The charge density from the proteins assumes that the
proteins have the mean charge of $-3$ that they would have if their
density and charge were uncorrelated.
\label{t2}
}
\end{table}


Future work could consider mixtures of proteins, ultimately aiming
to understand the cytosol of a living cell, which is a mixture
of of order $10^3$ different
types of proteins as well as DNA, RNA,
ions like ATP and potassium, etc.. This is of course very complex
but {\it if} in the cytosol
the proteins of {\em E. coli}
are present in amounts which are uncorrelated with their net charge,
the mean charge of the proteins will be close to the
mean of the distribution of Fig.~\ref{qcoli}. This is quite small.
Neidhardt \cite{neidhardt}
has taken an inventory of the species inside {\em E. coli}, and the
results for charged macromolecules are shown in Table~\ref{t2}.
The charged macromolecules in a cell are protein, DNA and the various
forms of RNA: transfer RNA (tRNA), messenger RNA (mRNA) and the
RNA in ribosomes (rRNA). See Refs.~\onlinecite{stryer,thecell}
for an introduction to the proteins, DNA and RNA.
Although for every molecule of tRNA molecule 
there are 10 of protein, for every ribosome there are 100 molecules
of protein, and  for every mRNA there are 1000 molecules of protein,
the contributions of the tRNA, ribosomes, DNA and proteins,
to the overall charge density of the macromolecules are
very roughly comparable. The ribosomes contribute the largest amount.
The macromolecules are negatively
charged and this negative charge is balanced by potassium ions \cite{thecell}.
Thus the cytosol resembles a solution of a negatively charged
polyelectrolyte, except that there is not one, relatively simple,
macromolecular species, but thousands of rather complex and diverse
species of macromolecules.


\section*{Acknowledgements}

It is a pleasure to acknowledge discussions with J. Cuesta, D. Frenkel
and P. Warren.


\begin{thebibliography}{99}
\bibitem{nomen} A brief introduction to the biological nomenclature:
an organism's DNA contains many genes, each of which codes for a
protein. The complete set of genes is called the organism's genome
and we will refer to the complete set of proteins as its proteome.
Some authors use the word proteome somewhat differently,
they use it to denote the set of proteins present
in the cytosol of
an organism at a particular time. See for example \cite{thecell}
for a more detailed definition of a genome.
\bibitem{aalys} The sequence used to determine this charge has
a Protein Data Bank ID \cite{pdb} of 1GWD.
\end{thebibliography}
\end{document}